\documentclass[aps,pra,twocolumn,groupedaddress]{revtex4}

\usepackage[dvips]{graphicx}
\usepackage{epsfig}
\usepackage{amssymb}
\usepackage{amsmath}
\usepackage{color}

\begin{document}


\title{Probing resonant energy transfer in collisions of ammonia with Rydberg helium atoms by microwave spectroscopy} 




\author{V. Zhelyazkova}
\altaddress{Present address: Laboratorium f\"ur Physikalische Chemie, ETH Z\"urich, CH-8093 Z\"urich, Switzerland}
\author{S. D. Hogan}
\affiliation{Department of Physics and Astronomy, University College London, Gower Street, London WC1E 6BT, U.K.}%




\begin{abstract}
We present the results of experiments demonstrating the spectroscopic detection of F\"{o}rster resonance energy transfer from NH$_3$ in the $X\,^1A_1$ ground electronic state to helium atoms in 1s$n$s\,$^3$S$_1$ Rydberg levels, where $n=37$ and $n=40$. For these values of $n$ the 1s$n$s\,$^3$S$_1\rightarrow$1s$n$p\,$^3$P$_J$ transitions in helium lie close to resonance with the ground-state inversion transitions in NH$_3$, and can be tuned through resonance using electric fields of less than 10~V/cm. In the experiments, energy transfer was detected by direct state-selective electric field ionization of the $^3$S$_1$ and $^3$P$_J$ Rydberg levels, and by monitoring the population of the $^3$D$_J$ levels following pulsed microwave transfer from the $^3$P$_J$ levels. Detection by microwave spectroscopic methods represents a highly state selective, low-background approach to probing the collisional energy transfer process and the environment in which the atom-molecule interactions occur. The experimentally observed electric-field dependence of the resonant energy transfer process, probed both by direct electric field ionization and by microwave transfer, agrees well with the results of calculations preformed using a simple theoretical model of the energy transfer process. For measurements performed in zero electric field with atoms prepared in the 1s40s\,$^3$S$_1$ level the transition from a regime in which a single energy transfer channel can be isolated for detection to one in which multiple collision channels begin to play a role has been identified as the NH$_3$ density was increased. 
\end{abstract}

\pacs{}

\maketitle 

\section{Introduction}\label{sec:intro}

Resonant energy transfer processes, in which electronic, vibrational or rotational energy is transferred from one atom or molecule to another, or between different parts of a single molecule, play important roles in a wide range of natural phenomena. The physical mechanism behind these processes was described by F\"{o}rster~\cite{forster48} in terms of excitation transfer between donor and acceptor molecules interacting by electric dipole-dipole interactions. The resonant interaction between such systems, $V_{\mathrm{dd}}$, can be expressed as~\cite{scholes03a}
\begin{eqnarray}
V_{\mathrm{dd}}(\vec{R})&=&\frac{1}{4\pi\epsilon_0}\Big[\frac{|\vec{\mu}_\mathrm{A}||\vec{\mu}_\mathrm{B}|}{R^3}-3\frac{(\vec{\mu}_{\mathrm{A}}\cdot\vec{R})(\vec{\mu}_{\mathrm{B}}\cdot\vec{R})}{R^5}\Big],
\label{eq1}
\end{eqnarray}
where $\vec{\mu}_{\mathrm{A}}$ and $\vec{\mu}_{\mathrm{B}}$ are the transition dipole moments in systems $A$ and $B$, respectively, and $R = |\vec{R}|$ is the inter-system distance.

These F\"orster resonance energy transfer processes are widely exploited in the condensed phase in spectroscopic approaches to measuring distances between fluorescent tags in proteins~\cite{latt65,stryer78}, in time-resolved mapping of conformation changing and folding, e.g., in RNA~\cite{millar01}, and play a role in energy transfer in light-harvesting  complexes~\cite{scholes03a}. In the gas phase, at high temperature or pressure these processes contribute to spectral line broadening~\cite{anderson49}, for example, in collisions of C$_2$H$_2$ with He which are of astrophysical interest and must be considered when modeling and interpreting spectra of planetary atmospheres~\cite{heijmen99}. In collisions of cold molecules at low energies we foresee opportunities to exploit the long-range resonant dipole-dipole interactions that lead to energy transfer between reactants to regulate access to short-range chemical processes including, for example, Penning ionization~\cite{hotop70a,henson12a,jankunas15a}. It has also been recently suggested that the resonant transfer of energy between Rydberg atoms and cold polar molecules could be used to efficiently and non-destructively detect the molecules~\cite{zeppenfeld17a}, and to characterize their rotational state populations~\cite{kuznetsova16a}. The electric dipole-dipole interactions associated with F\"orster resonance energy transfer have also be considered in schemes for coherent control~\cite{kuznetsova11a} and molecular cooling~\cite{huber12,zhao12}.

Atoms or molecules in highly excited Rydberg states are particularly well suited to studies of resonant energy transfer in the gas phase~\cite{safinya81,gallagher92a,vogt06a,gallagher08a,gunter13a,ravets14}. This is because of the wide range of Rydberg-Rydberg transition frequencies that can be chosen from to ensure that the resonance condition with a particular collision partner is satisfied, and the large electric dipole transition moments (on the order of 1000~D for $n\geq30$) associated with these transitions. For these reasons, a number of studies have been performed in the past to probe resonant energy transfer between Rydberg atoms and thermal samples of polar ground state molecules. In the absence of applied external fields, measurements of the discrete transfer of rotational energy from NH$_3$ to the electronic degrees of freedom of Xe Rydberg atoms~\cite{smith78}, studies of the effects of collisions with NH$_3$\,~\cite{petitjean86} and CO\,~\cite{petitjean84} on the fluorescence lifetimes of Rydberg states of Rb, and the observation of contributions from the inversion transitions in NH$_3$ and ND$_3$ to the ionization of K atoms in Rydberg states with $n=150$\,~\cite{ling93} have been performed. Most recently, we have demonstrated control over the resonant transfer of energy from the inversion sublevels in NH$_3$ to Rydberg He atoms for the first time~\cite{zhelyazkova17a}.

In the experiments reported here, we extend our recent work on collisions of NH$_3$ and Rydberg He atoms with further studies of the dependence of the energy transfer process on the strength of applied dc electric fields, and on the density of NH$_3$. In addition to the method of Rydberg-state-selective electric field ionization that we employed previously to identify state changing arising as a result of resonant energy transfer, we have now also implemented microwave spectroscopic techniques to probe the energy transfer process and the environment in which the atom-molecule interactions occur. In the experiments, an effusive beam of NH$_3$, in the X\,$^1$A$_1$ ground electronic state interacted with He atoms excited to 1s$n$s\;$^3$S$_1\equiv |n\mathrm{s}\rangle$ levels, with $n=37$ and $n=40$, in a pulsed supersonic beam. At the 300~K operating temperature of the NH$_3$ source the inversion transitions in the most-populated lowest vibrational level of the X\,$^1$A$_1$ state lie at $\Delta E/h\simeq23$~GHz  ($\Delta E/hc\simeq0.79$~cm$^{-1}$), and are close to resonant with the $|n\mathrm{s}\rangle\rightarrow |n\mathrm{p}\rangle$ transitions (1s$n$p\;$^3$P$_J\equiv |n\mathrm{p}\rangle$) in He for $n=36-41$. The collisional energy transfer process studied is therefore
\begin{eqnarray}
\mathrm{NH}_3\,(\mathrm{X}\,^1\mathrm{A}_1,v=0,J,K,-)\; +\; \mathrm{He}\,(1\mathrm{s}n\mathrm{s}\,^3\mathrm{S}_1) &\rightarrow&\nonumber\\
& & \hspace*{-6.6cm}\mathrm{NH}_3\,(\mathrm{X}\,^1\mathrm{A}_1,v=0,J,K,+)\; +\; \mathrm{He}\,(1\mathrm{s}n\mathrm{p}\,^3\mathrm{P}_J)
\label{eq2}
\end{eqnarray}
where $-$($+$) indicates the upper anti-symmetric (lower symmetric) inversion sublevel, $v$ is the vibrational quantum number, and $J$ and $K$ are the total angular momentum quantum number, and the projection of $\vec{J}$ onto the symmetry axis of the molecule, respectively.

The remainder of this paper is structured as follows: In Sec.~\ref{sec:theory} the theoretical model used to describe the resonant energy transfer processes is presented before the apparatus used in the experiments is described in Sec.~\ref{sec:expt}. In Sec.~\ref{sec:results1} the results of experiments performed with atoms initially photoexcited to the $|37\mathrm{s}\rangle$ Rydberg state, and which involved detection by direct state-selective electric field ionization, as used previously in the work reported in Ref.~\cite{zhelyazkova17a}, and microwave spectroscopic methods, are presented. These include the demonstration of electric-field controlled F\"orster resonance energy transfer. The dependence of the energy transfer process on the NH$_3$ density in zero electric field for atoms initially prepared in the $|40\mathrm{s}\rangle$ state is then presented along with microwave spectroscopic studies of Rydberg-Rydberg transitions in atoms after undergoing energy transfer. Finally, in Sec.~\ref{sec:conc} conclusions are drawn and areas for future work are briefly outlined.

\section{Theoretical background}\label{sec:theory}

The process of resonant energy transfer from NH$_3$ to Rydberg He atoms studied here can be described theoretically using an approach similar to that employed to treat energy transfer between pairs of Rydberg atoms in Ref.~\cite{gallagher92a}. In the case of interest here, the electric dipole transition moments of the collision partners, $\vec{\mu}_{\mathrm{A}}$ and $\vec{\mu}_{\mathrm{B}}$ in Eq.~\ref{eq1}, are replaced by $\vec{\mu}_{\mathrm{He}}=\langle n\mathrm{p}|e\,\hat{\vec{r}}\,|n\mathrm{s}\rangle$ and $\vec{\mu}_{\mathrm{NH}_3}=\langle +|e\,\hat{\vec{r}}\,|-\rangle$, respectively. Because of the angular dependence of the second term in the square brackets in Eq.~\ref{eq1}, the absolute value of $V_{\mathrm{dd}}$ ranges from 0 to $2\mu_{\mathrm{NH}_3}\mu_{\mathrm{He}}/(4\pi\epsilon_0 R^3)$, depending on the relative orientation of the dipole moments. In the experiments described here, when the atoms and molecules approached each other their dipole moments were randomly oriented with respect to each other. With this in mind, and without considering effects of dynamical orientation as the dipoles approach each other, the typical interaction potential, $\tilde{V}_{\mathrm{dd}}$, is simply assumed to be
\begin{eqnarray}
\tilde{V}_{\mathrm{dd}}(R) &=& \frac{\mu_{\mathrm{He}}\mu_{\mathrm{NH}_3}}{4\pi\epsilon_0R^3}. 
\label{eq3}
\end{eqnarray}
The $R^{-3}$ dependence of $\tilde{V}_{\mathrm{dd}}$ permits the assumption that
\begin{eqnarray}
\tilde{V}_{\mathrm{dd}} &=& \frac{\mu_{\mathrm{He}}\mu_{\mathrm{NH}_3}}{4\pi\epsilon_0b^3}\hspace*{0.8cm}\mathrm{for}\hspace{0.5cm} R\leq b\nonumber\\
\tilde{V}_{\mathrm{dd}}   &=& 0 \hspace{1.75cm}\mathrm{for}\hspace{0.5cm} R>b
\label{eq4}
\end{eqnarray}
where $b$ is the impact parameter. Considering an atom-molecule interaction time, $t$, during which $R\leq b$, energy transfer is assumed to occur when $\tilde{V}_{\mathrm{dd}}t/h\gtrsim1$, i.e., the product of the interaction rate and interaction time exceed 1. For a relative collision speed $v$, the interaction time is $t\simeq b/v$ when $\tilde{V}_{\mathrm{dd}}t/h=1$. Since the cross-section associated with a single scattering channel with an impact parameter $b$ is $\sigma=\pi b^2$, the cross-section for resonant energy transfer, $\sigma_\mathrm{R}$, is therefore
\begin{eqnarray}
\sigma_{\mathrm{R}}&\simeq& \frac{\mu_{\mathrm{He}}\mu_{\mathrm{NH}_3}}{4\epsilon_0vh}.
\label{eq5}
\end{eqnarray}

In accounting for the dependence of $\mu_{\mathrm{He}}$ on the interaction electric field, $F_{\mathrm{int}}$, the Rydberg states that evolve adiabatically to the pure $|n\mathrm{s}\rangle$ and $|n\mathrm{p}\rangle$ levels in zero field are denoted $|n\mathrm{s}'\rangle$ and $|n\mathrm{p}'\rangle$, respectively. The corresponding electric dipole transition moments are then $|\langle n\mathrm{p}'|\hat{\mu}|n\mathrm{s}'\rangle| = \mu_{n\mathrm{s}',n\mathrm{p}'}(F_{\mathrm{int}})$. The electric dipole transition moments associated with the inversion of the NH$_3$ molecule, $\mu_{\mathrm{NH}_3}=\mu_{J,K}=|\langle J,K,+|\hat{\mu}|J,K,-\rangle|$, are unaffected by the relatively weak electric fields used in the experiments. However, they do depend on the rotational quantum numbers of the states populated. This leads to a set of transition dipole moments~\cite{townes55a}
\begin{eqnarray}
\mu_{J,K} &=& \sqrt{\frac{\mu_0^2K^2}{J(J+1)}},
\label{eq7}
\end{eqnarray}
where $\mu_0=1.468$~D.

\begin{figure}
\includegraphics[clip=,width=7.5cm]{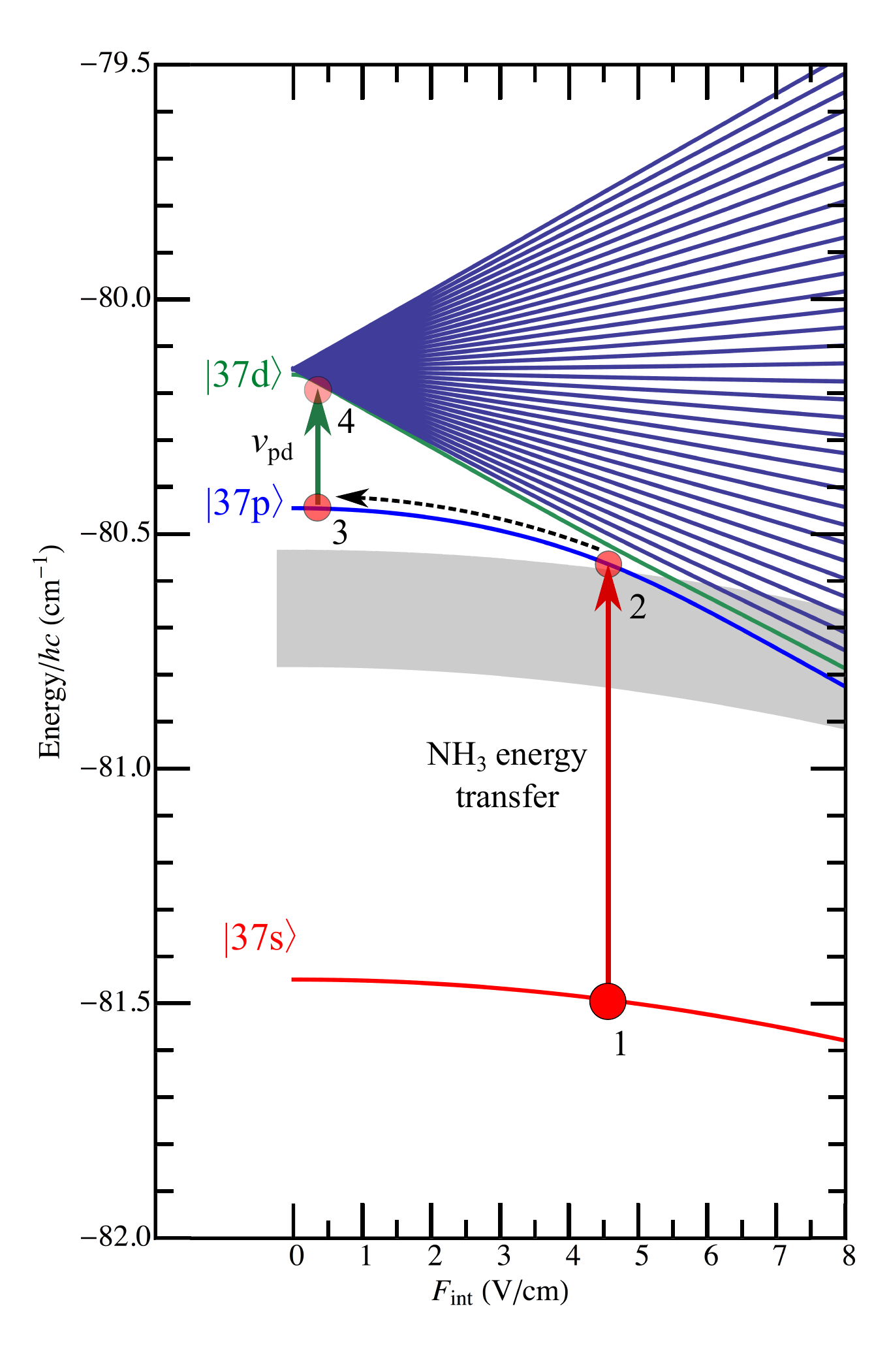}
\caption{\label{fig1} Calculated Stark map of the manifold of triplet Rydberg states in He with $n=37$. The center of the shaded gray band corresponds to the energy of the $|37\mathrm{s}'\rangle$ state offset by the centroid inversion splitting in NH$_3$ at 300~K ($\sim0.79$~cm$^{-1}$). The width of this band reflects the FWHM of the set of inversion transitions of $0.125$~cm$^{-1}$. The red point labelled 1 represents a typical position in the Stark map in which collisions with NH$_3$ leads to energy transfer and excitation to the $|37\mathrm{p}\rangle$ state (red vertical arrow from 1 to 2). After an interaction time of 5~$\mu$s, $F_{\mathrm{int}}$ is reduced to zero (dashed black curve from 2 to 3) and a pulse of microwave radiation, $\nu_{\mathrm{pd}}$, resonant with the $|37\mathrm{p}\rangle\rightarrow|37\mathrm{d}\rangle$ transition in zero electric field is applied (green vertical arrow from 3 to 4).}
\end{figure}

In zero electric field the $|40\mathrm{s}\rangle\rightarrow|40\mathrm{p}\rangle$ transition which occurs in He at 0.794~cm$^{-1}$ ($\equiv23.80$~GHz) is approximately resonant with the spectral-intensity-weighted centroid inversion transition frequency in a room temperature gas of NH$_3$, 0.78~cm$^{-1}$ ($\equiv23.38$~GHz). In this case, $|\langle 40\mathrm{p}|\hat{\mu}|40\mathrm{s}\rangle|\simeq$~3300~D and, for a relative collision speed of $v=2000$~m/s, which is typical under the conditions of the experiments described here (see Sec.~\ref{sec:expt}), a cross-section for resonant energy transfer of $\sigma_{\mathrm{R}}=1.1\times10^{-11}$~cm$^{2}$ results. 

For the triplet Rydberg states in He with values of $n<40$ the $|n\mathrm{s}\rangle\rightarrow|n\mathrm{p}\rangle$ transition in zero electric field lies higher than the centroid room-temperature inversion transition frequency. This can be seen in Fig.~\ref{fig1} for the case in which $n=37$. This figure contains a Stark map of the triplet $n=37$ manifold of Rydberg states in He for electric fields up to 8~V/cm. When calculating this Stark map the quantum defects, $\delta_{n\ell}$, of the states with low electron orbital angular momentum, $\ell$, were obtained from the results reported in Ref.~\cite{drake99}, i.e., $\delta_{37\mathrm{s}}=0.296685$, $\delta_{37\mathrm{p}}=0.068347$ and $\delta_{37\mathrm{d}}=0.002887$, and $\delta_{37\mathrm{f}}=0.000446$. 

\begin{figure}
\includegraphics[clip=,width=7.5cm]{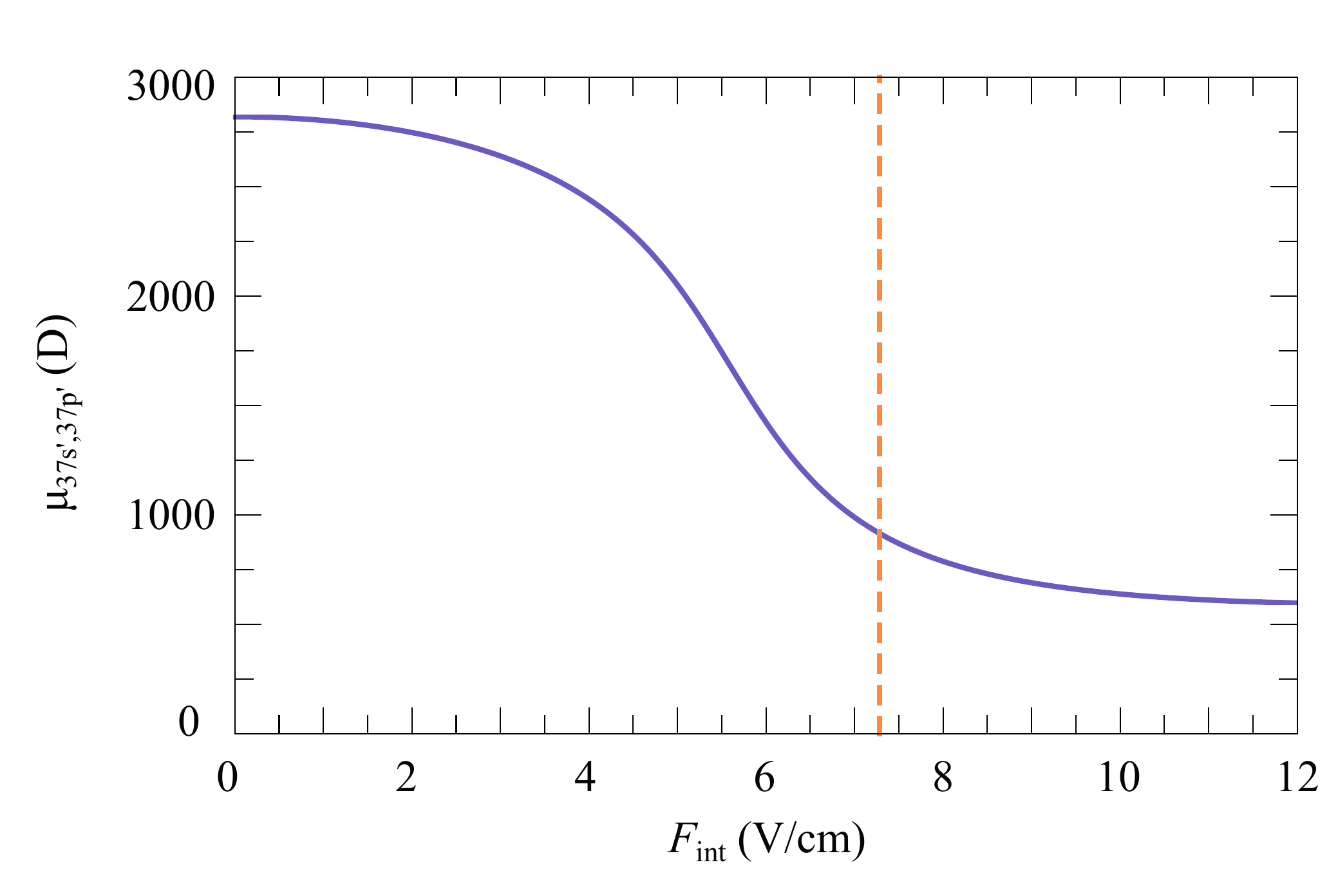}
\caption{\label{fig2} The dependence of the electric dipole transition moment, $\mu_{37\mathrm{s}',37\mathrm{p}'}$, in He on the electric field strength, $F_{\mathrm{int}}$. The dashed vertical line indicates the value of $F_{\mathrm{int}}$ for which the $|37\mathrm{s}'\rangle \rightarrow |37\mathrm{p}'\rangle$ transition is resonant with the $|1,1,-\rangle\rightarrow|1,1,+\rangle$ inversion transition in NH$_3$.}
\end{figure}

In zero electric field the $|37\mathrm{s}\rangle$, $|37\mathrm{p}\rangle$ and $|37\mathrm{d}\rangle$ states lie lower in energy than the higher angular momentum states. For this value of $n$, the zero-field $|n\mathrm{s}\rangle\rightarrow|n\mathrm{p}\rangle$ transition lies above the centroid inversion transition frequency in NH$_3$. The gray shaded band, the center of which is at $\sim-80.65$~cm$^{-1}$ in zero field, in the figure corresponds to the energy of the $|37\mathrm{s}\rangle$ state added to which is the centroid inversion transition wavenumber in a room temperature sample of NH$_3$. The width of this band represents the full-width-at-half-maximum (FWHM) of the set of spectral-intensity-weighted inversion transition wavenumbers. As can be seen the process of resonant energy transfer between the two systems can be tuned into resonance upon the application of appropriate electric fields. In these situations the transition dipole moments, and hence energy transfer cross sections, depend on the strength of the applied electric field. Transition dipole moments, $\mu_{37\mathrm{s}',37\mathrm{p}'}$, calculated from the eigenvectors of the complete Hamiltonian matrix describing the interaction of the Rydberg atom with an external electric field~\cite{zimmerman79a}, are displayed in Fig.~\ref{fig2} for fields up to 12~V/cm. The electric field, $F_{\mathrm{int}}$, for which the $|37\mathrm{s}'\rangle\rightarrow |37\mathrm{p}'\rangle$ transition in He is resonant with the $|1,1,-\rangle\rightarrow|1,1,+\rangle$ inversion transition in NH$_3$ is indicated by the dashed vertical line in Fig.~\ref{fig2}. The calculated transition dipole moment $\mu_{37\mathrm{s}',37\mathrm{p}'}$ has a maximum value of $\sim2800$~D in zero field and decreases as the electric field increases. This dependence of the transition dipole moment on the electric field strength shows a similar general trend for each value of $n$ in the range from 36 to~40.

From the data in Fig.~\ref{fig1} and in Fig.~\ref{fig2}, the electric field dependence of the energy transfer process, which arises through the Rydberg state transition dipole moments $\mu_{n\mathrm{s}',n\mathrm{p}'}(F_{\mathrm{int}})$, and the detuning from resonance of the inversion transition for each rotational state populated in the NH$_3$ beam, is determined. To calculate a quantity which is proportional to the signal measured in the experiments a Gaussian dependence of the energy transfer rate on the detuning from resonance is assumed. The FWHM of this Gaussian function, $\Delta E_{\mathrm{FWHM}}$, is derived from the inverse of the atom-molecule interaction time to be
\begin{eqnarray}
\frac{\Delta E_{\mathrm{FWHM}}}{h} \simeq \frac{v}{b} &=& \sqrt{ \frac{4\pi\epsilon_0 v^3 h}{\mu_{\mathrm{He}} \mu_{\mathrm{NH}_3} }}\\\nonumber\\
&=& \frac{C_{\mathrm{width}}}{ \sqrt{\mu_{\mathrm{He}} \mu_{\mathrm{NH}_3}}}.\label{eq:width} 
\end{eqnarray}
In using this model, which was also employed in Ref.~\cite{zhelyazkova17a}, to aid in the interpretation of the experimental data only one fit parameter is required to allow comparison with the observations -- the constant of proportionality associated with the resonance width, $C_{\mathrm{width}}$. 
    
\begin{figure}
\includegraphics[clip=,width=7.5cm]{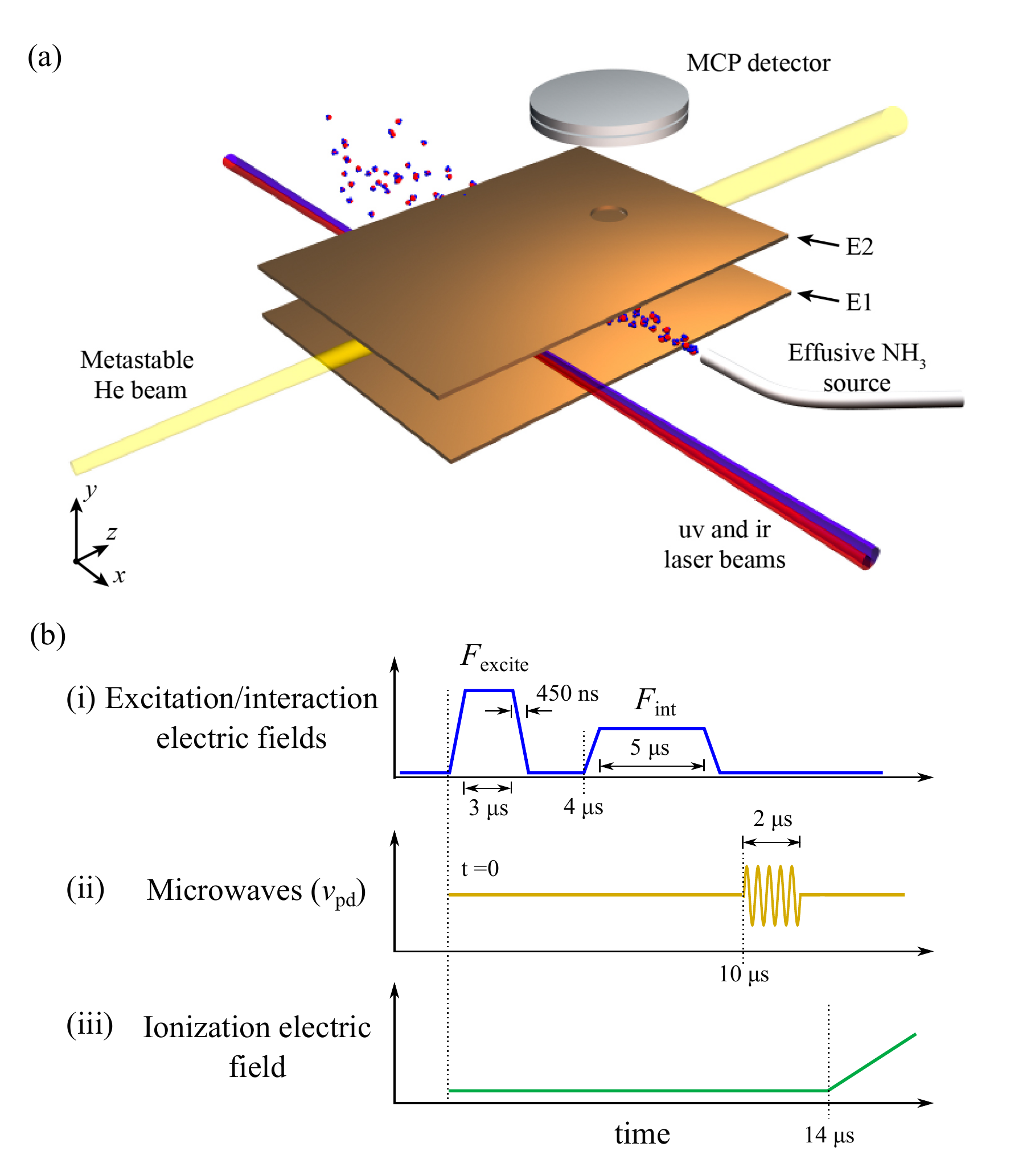}
\caption{\label{fig3} (a) Schematic diagram of the experimental apparatus. (b) Typical timing sequence indicating each phase of a single experimental cycle.}
\end{figure}

\section{Experiment}\label{sec:expt}

The apparatus used in the experiments is similar to that described in Ref.~\cite{zhelyazkova17a}. A pulsed supersonic beam of He atoms in the metastable $1\mathrm{s}2\mathrm{s}\,^3\mathrm{S}_1$ level was generated in a dc electric discharge at the exit of a pulsed valve. The discharge was seeded with electrons from a heated tungsten filament~\cite{halfmann00} and the valve was operated at a repetition rate of 50~Hz. After passing through a 2-mm-diameter skimmer, and an electrostatic filter to remove stray ions produced in the discharge, the atomic beam crossed two co-propagating continuous wave (cw) laser beams. The first laser beam was frequency stabilized to be resonant with the $1\mathrm{s}2\mathrm{s}\,^3\mathrm{S}_1 \rightarrow 1\mathrm{s}3\mathrm{p}\,^3\mathrm{P}_2$ transition in He at 25708.5876~cm$^{-1}$ ($\equiv388.9751$~nm), and the second was tuned to drive $1\mathrm{s}3\mathrm{p}\,^3\mathrm{P}_2\rightarrow 1\mathrm{s}n\mathrm{s}\,^3\mathrm{S}_1$ transitions in the presence of a weak electric field, where $n=37$ or $n=40$. These transitions occur at 12664.656~cm$^{-1}$ ($\equiv789.599$~nm) for $n=37$, and 12676.499~cm$^{-1}$ ($\equiv788.861$~nm) for $n=40$. The laser beams were focused to FWHM beam waists of $\sim$100~$\mu$m. The longitudinal speed of the ensemble of excited Rydberg atoms was $\sim2000$~m/s. The longitudinal translational temperature of the atomic beam in the moving frame of reference was $\sim0.5$~K, and the transverse temperatures were less than $10$~mK. The density of excited atoms was estimated to be on the order of $10^7$~cm$^{-3}$ (see Ref.~\cite{zhelyazkova16a}).

As indicated in the schematic diagram in Fig.~\ref{fig3}(a) Rydberg state photoexcitation took place between a pair of 70~mm$\times$100~mm,  copper plates labelled E1 and E2. To prepare a spatially localized ensemble of Rydberg atoms, photoexcitation of the $|37\mathrm{s}'\rangle$ or $|40\mathrm{s}'\rangle$ states was performed in the presence of a pulsed electric field, $F_{\mathrm{excite}}$, for a time of 3~$\mu$s [see Fig.~\ref{fig3}(b)]. This field was generated by applying a pulsed potential to electrode E2 to bring the atomic transition into resonance with the infrared laser which was detuned slightly below the $1\mathrm{s}3\mathrm{p}\,^3\mathrm{P}_2\rightarrow 1\mathrm{s}n\mathrm{s}\,^3\mathrm{S}_1$ transition wavenumber in zero electric field. In the experiments performed with atoms excited to the $|37\mathrm{s}\rangle$ state, $1$~$\mu$s after $F_{\mathrm{excite}}$ was switched to zero, a second adjustable, pulsed electric potential was applied to E2 for a time of $5$~$\mu$s [see Fig.~\ref{fig3}(b-i)]. This pulsed potential was chosen to generate selected interaction electric fields, $F_{\mathrm{int}}$, in which the collisions with the NH$_3$ occurred. 

The NH$_3$ molecules were introduced into the interaction region of the apparatus through a 1-mm-diameter flexible stainless steel tube. This effusive molecular-beam source was operated at room temperature. The pressure at which the NH$_3$ source was maintained was measured using a Pirani gauge and was adjusted in the range from $P_{\mathrm{NH}_3}=0.01$ to 1.8~mbar. Under normal operating conditions the pressure in the main vacuum chamber was $\lesssim10^{-7}$~mbar and rose to $\sim10^{-6}$~mbar when the ammonia was present. The mean speed of the beam of NH$_3$ molecules in the interaction region of the apparatus was calculated to be 720~m/s\,~\cite{scholes88a}, with an expected number density of $\sim10^{9}$~cm$^{-3}$ for $P_{\mathrm{NH}_3}=1$~mbar. 

After the atoms interacted with the ammonia molecules for $5~\mu$s, a microwave pulse with a duration of $\sim2~\mu$s and a frequency, $\nu_{\mathrm{pd}}$, resonant with the $|37\mathrm{p}\rangle\rightarrow|37\mathrm{d}\rangle$ (or $|40\mathrm{p}\rangle\rightarrow|40\mathrm{d}\rangle$) transition in zero electric field could then be applied [Fig.~\ref{fig3}(b-ii)]. The microwave radiation was coupled into the interaction region in the apparatus from a copper antenna located outside a 15~mm-diameter quartz vacuum window. Following the interaction of the atoms with the microwave field, and $\sim12.5~\mu$s after laser photoexcitation, a slowly-rising time-dependent electric potential was applied to the copper plate E1, to generate a time-dependent electric field that increased at a rate of $\sim300$~V/(cm\,$\mu$s) to allow state-selective detection of the Rydberg atoms by electric field ionization. The ionized electrons were accelerated through a hole in plate E2 and toward a microchannel plate (MCP) detector. This detection region of the apparatus was located $\sim25$~mm downstream from the position of Rydberg state photoexcitation. The radiative lifetime of, e.g., the $|37\mathrm{s}\rangle$ state in He is $\sim40$~$\mu$s. Since this is significantly longer than the flight time of the Rydberg atoms from their position of photoexcitation to the detection region ($\sim12.5$~$\mu$s) changes in the lifetime of this state in the time-dependent electric fields used in the experiment did not have a significant effect on the measurements.

\section{Results}\label{sec:results1}

F\"orster resonance energy transfer in collisions of NH$_3$ with Rydberg He atoms was studied for two different initially prepared Rydberg states, the $|37\mathrm{s}\rangle$ and $|40\mathrm{s}\rangle$ states. In zero electric field the $|37\mathrm{s}\rangle\rightarrow|37\mathrm{p}\rangle$ transition lies above the centroid inversion transition frequency in NH$_3$ but can be tuned into resonance with the inversion transitions using electric fields between $5$ and $8$~V/cm. On the other hand, the $|40\mathrm{s}\rangle\rightarrow|40\mathrm{p}\rangle$ transition is approximately resonant with the centroid inversion transition frequency in zero field. 

\begin{figure}
\begin{center}
\includegraphics[clip=,width=8cm]{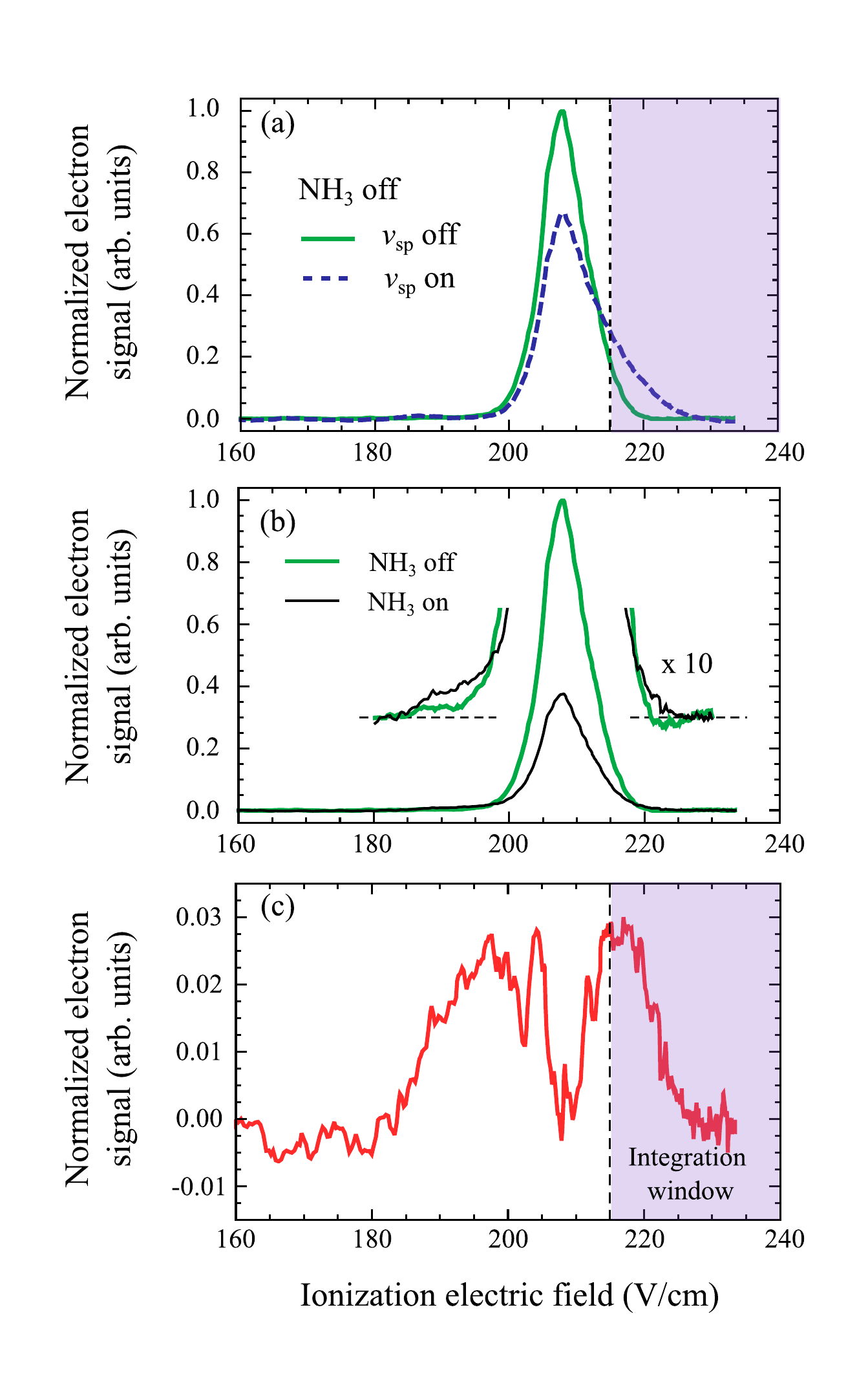}
\caption{\label{fig4} Electron signals recorded following direct electric field ionization of Rydberg atoms initially prepared in the $|37\mathrm{s}\rangle$ state [thick green curve in (a) and (b)]. The dashed blue curve in (a) results from the transfer of population to the $|37\mathrm{p}\rangle$ state by a pulse of resonant microwave radiation at a frequency $\nu_{\mathrm{sp}}$. Data recorded in the absence of microwave radiation but following collisions with NH$_3$ is presented as the thin black curve in (b). An expanded view of the data in (b) with the NH$_3$ beam on and off is also included with a vertical offset of 0.3. The difference between the normalized ionization profile with the NH$_3$ present and that recorded in the absence of the NH$_3$ is displayed in (c). The shaded region between ionization fields of 215 and 240 V/cm in (c) corresponds to the integration window referred to in the text.}
\end{center}
\end{figure}

\subsection{Detection by direct state-selective electric field ionization }

Considering first the measurements preformed at $n=37$. The experimental data displayed in Fig.~\ref{fig4} demonstrate the effect of the energy transfer process on the electron signal recorded following direct ionization of the excited Rydberg atoms in a slowly-rising time-dependent electric field without applying the microwave pulse in Fig.~\ref{fig3}(b-ii). In Fig.~\ref{fig4}(a) the dependence of the electron signal resulting from ionization of the $|37\mathrm{s}\rangle$ state in the absence of NH$_3$ is displayed as the continuous green curve. The intensity maximum of this signal occurs for an ionization field of $\sim208$~V/cm. To obtain a reference for how the electric-field-ionization signal changes when population is transferred to the $|37\mathrm{p}\rangle$ state, a pulse of microwave radiation, $\nu_{\mathrm{sp}}$, resonant with the $|37\mathrm{s}\rangle\rightarrow|37\mathrm{p}\rangle$ transition in zero field, was applied between the time of laser photoexcitation and pulsed electric field ionization. The resulting data is displayed as the dashed blue curve in Fig.~\ref{fig4}(a) and shows an increase in signal for high ionization fields, i.e., for fields greater than 215~V/cm. These higher fields required to ionise the $|37\mathrm{p}\rangle$ state are commensurate with transfer of population to $|m_{\ell}|=1$ sublevels which ionise quasi-diabatically. 

After recording the reference data by selective preparation of the initial and final Rydberg states of interest, the microwave source was turned off and replaced with the effusive molecular beam. The electric field ionization data recorded with this beam operated at $P_{\mathrm{NH}_3}=1.8$~mbar and for an interaction field of $F_{\mathrm{int}}=3.6$~V/cm is displayed as the thin black curve in Fig.~\ref{fig4}(b). The overall reduction in intensity with respect to the data recorded in the absence of the NH$_3$ is a result of  Penning ionization, and ionizing and Rydberg-state-changing collisions in which rotational energy, and the energy associated with the inversion transitions, is transferred from the molecules. However, a small enhancement in the signal at and around the ionization field of the $|37\mathrm{p}\rangle$ state can be seen. This increase in the $|37\mathrm{p}\rangle$ signal can be isolated by normalizing the data recorded with the NH$_3$ present so that the intensity maxima of the two data sets at $\sim208$~V/cm are equal. The contribution to the signal from the $|37\mathrm{s}\rangle$ state recorded in the absence of NH$_3$ is then subtracted from the signal obtained in the presence of the NH$_3$ to yield the difference signal displayed in Fig.~\ref{fig4}(c). This data shows a positive change in the signal intensity for low ionization fields, i.e., fields between 180 and 210~V/cm, which is attributed to effects of rotational energy transfer from the NH$_3$ molecules to Rydberg states with $n>37$. However, the data in Fig.~\ref{fig4}(c) also exhibit a positive change in signal intensity for ionization fields close to, and above, 215~V/cm where the $|37\mathrm{p}\rangle$ state ionizes. Using this measurement procedure a lower limit on the $|37\mathrm{p}\rangle$ electron signal arising as a result of resonant energy transfer in the atom-molecule collisions was obtained by integrating the difference signal in Fig.~\ref{fig4}(c) for ionization fields from 215 to 240~V/cm, as indicated. 

\subsection{Microwave assisted state-selective electric field ionization}

Further information on the process of resonant energy transfer was obtained by microwave spectroscopy of $|n\mathrm{p}\rangle\rightarrow |n\mathrm{d}\rangle$ transitions in zero electric field that can only occur following the collisional population of $|n\mathrm{p}\rangle$ Rydberg states. In carrying out these measurements ionization of the Rydberg atoms using time-dependent electric fields was also employed to permit state-selective detection, however, the addition of a pulse of microwave radiation to transfer atoms from the $|n\mathrm{p}\rangle$ to $|n\mathrm{d}\rangle$ Rydberg states at the end of the atom-molecule interaction period [see Fig.~\ref{fig3}(b)-ii], provided an unambiguous signature of the resonant energy transfer process in a single measurement. 

The effects of each of the electric field and microwave pulses, employed in the implementation of this detection scheme, on the Rydberg states populated can be seen from the numbered points in Fig.~\ref{fig1}. In the experiments, after initial photoexcitation of the $|37\mathrm{s}\rangle$ state, the excited atoms were polarized by applying an electric field $F_{\mathrm{int}}=4.9$~V/cm (red point labelled 1 in Fig.~\ref{fig1}). If this field is tuned to bring the $|37\mathrm{s}'\rangle\rightarrow|37\mathrm{p}'\rangle$ transition into resonance with the $|J,K,-\rangle\rightarrow|J,K,+\rangle$ inversion transitions, population transfer to the $|37\mathrm{p}'\rangle$ state (red vertical arrow from 1 to 2 in Fig.~\ref{fig1}) can occur. After a fixed atom-molecule interaction time the electric field $F_{\mathrm{int}}$ is decreased to zero (dashed black arrow from 2 to 3 in Fig.~\ref{fig1}) before a 2-$\mu$s-long pulse of microwave radiation, resonant with the $|37\mathrm{p}\rangle\rightarrow|37\mathrm{d}\rangle$ transition at a frequency of $\nu_{\mathrm{pd}} = 8.526$~GHz, is applied to transfer Rydberg atoms in the $|37\mathrm{p}\rangle$ state that have undergone resonant energy transfer with the NH$_3$, into the $|37\mathrm{d}\rangle$ state (green vertical arrow from 3 to 4 in Fig.~\ref{fig1}). A time-dependent ionization electric field is subsequently applied permitting state-selective detection of the atoms. 

\begin{figure}
\begin{center}
\includegraphics[clip=,width=8.0cm]{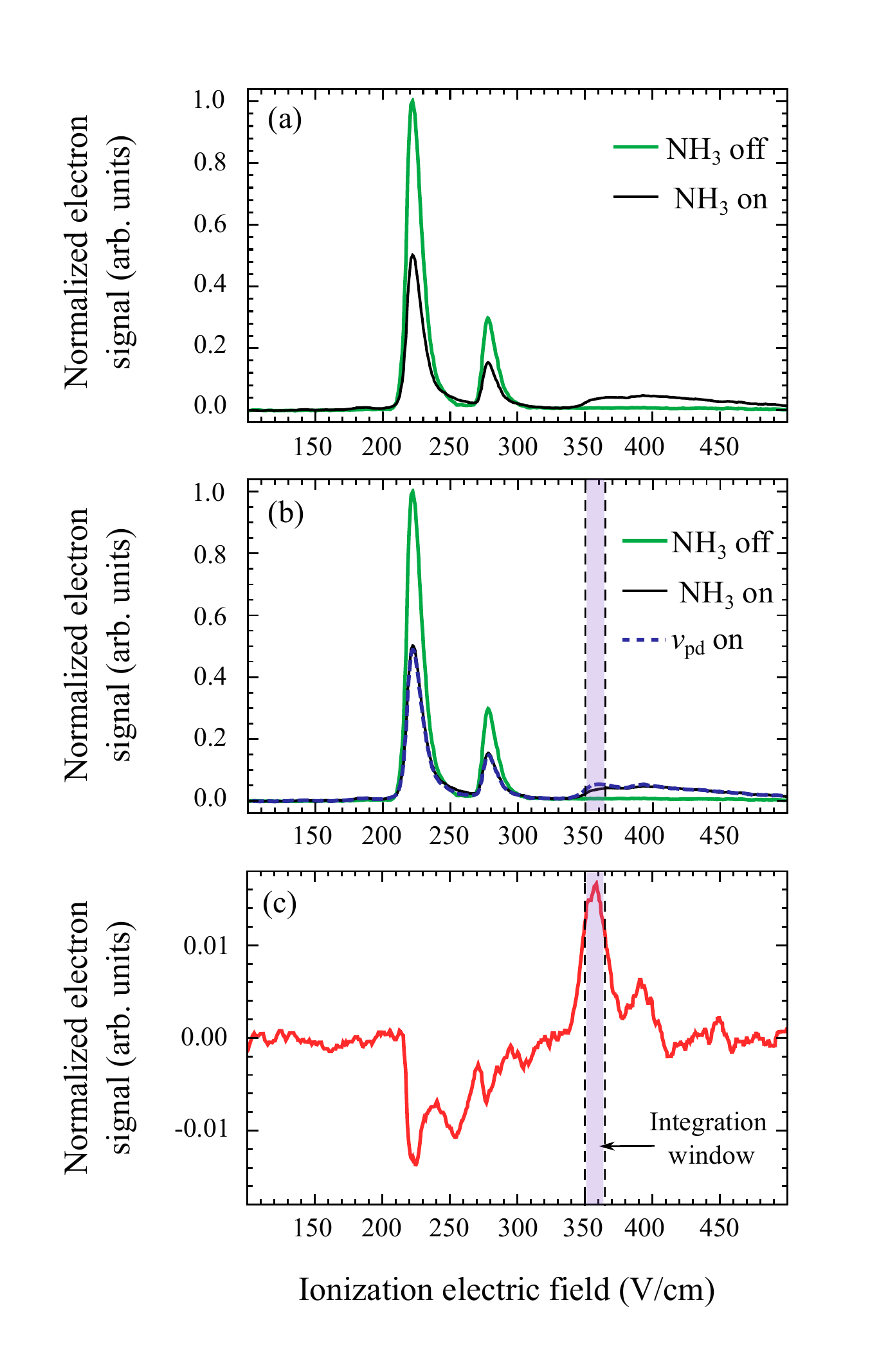}
\caption{\label{fig5} 
Electron signals recorded by electric field ionization of the Rydberg atoms after microwave transfer to the $|37\mathrm{d}\rangle$ state. Data recorded by direct electric field ionization with and without NH$_3$ when $F_{\mathrm{int}}=4.9$~V/cm are presented in (a). The addition of a pulse of microwave radiation to drive the $|37\mathrm{p}\rangle\rightarrow|37\mathrm{d}\rangle$ transition in atoms having undergone resonant energy transfer resulted in the data presented as the dashed blue curve in (b). The difference between the measurements with and without this microwave pulse, but with the NH$_3$ beam on, is displayed in (c). The shaded region between ionization fields of 350 and 360~V/cm corresponds to the integration window referred to in the text.}
\end{center}
\end{figure}

Using this detection scheme the ionization profiles displayed in Fig.~\ref{fig5} were recorded. In carrying out these measurements, a wider range of more rapidly changing ionization fields were employed than those used in recording the data in Fig.~\ref{fig4}. With only the $|37\mathrm{s}\rangle$ state populated the data presented as the continuous green curve in Fig.~\ref{fig5}(a) was recorded. The sharp features at ionization fields of $\sim220$~V/cm and $\sim280$~V/cm in this figure both originate from atoms in the $|37\mathrm{s}\rangle$ state. The differences in the structure of this signal from that in Fig.~\ref{fig4}(a) arise as a result of the time-dependence of the ionization field used which was optimized to ultimately detect atoms in the $|37\mathrm{d}\rangle$ state. These differences modify the electric field ionization dynamics at the avoided crossings in the Stark map for the Rydberg states, and the electron trajectories to the MCP detector in the apparatus. 

Upon introducing NH$_3$ into the apparatus at a source pressure of $P_{\mathrm{NH}_3}=0.9$~mbar, the electric field ionization signal displayed as the black curve in Fig.~\ref{fig5}(a) was recorded. Differences are clearly seen upon comparing this set of data and that recorded without NH$_3$, particularly for ionization fields above 300~V/cm. To isolate the signal associated with the energy transfer channel resonant with the inversion transitions in NH$_3$, the 2-$\mu$s-long pulse of microwave radiation, $\nu_{\mathrm{pd}}$, resonant with the $|37\mathrm{p}\rangle\rightarrow |37\mathrm{d}\rangle$ transition was applied at the end of the interaction period, prior to electric field ionization. The ionization signal resulting from this additional step in the experimental procedure is displayed as the dashed blue curve in Fig.~\ref{fig5}(b) where the data in Fig.~\ref{fig5}(a) is also included for direct comparison. Upon subtracting the signal recorded with $\nu_{\mathrm{pd}}$ off, from that recorded with $\nu_{\mathrm{pd}}$ on, the data presented in Fig.~\ref{fig5}(c) was obtained. The enhancement in the $|37\mathrm{d}\rangle$ electron signal at an ionization field of $\sim355$~V/cm can be clearly identified. This enhancement does not appear in the absence of the NH$_3$.

\begin{figure}
\includegraphics[clip=,width=7.5cm]{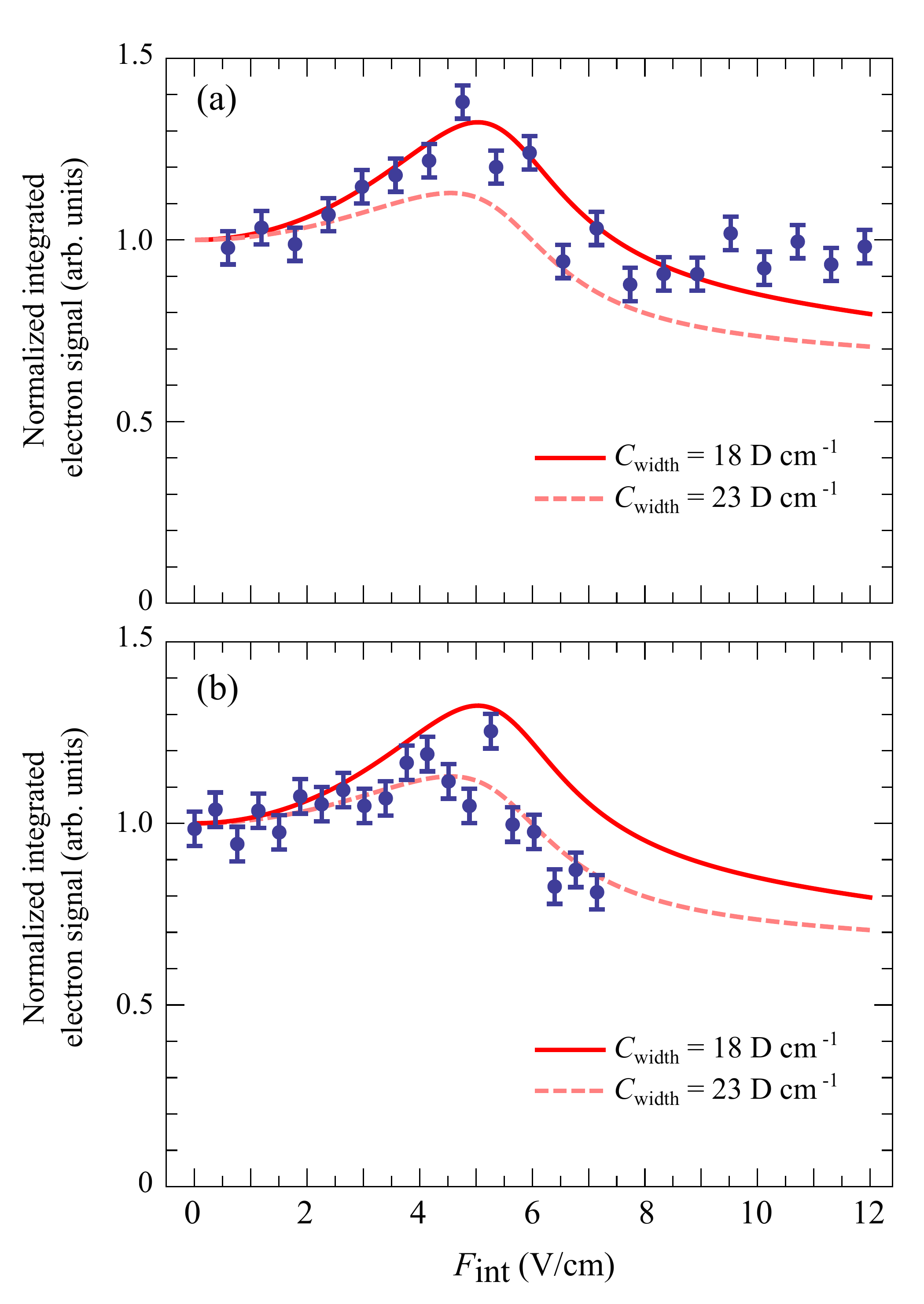}
\caption{The dependence of the integrated electron signal associated with the resonant transfer of energy from the $|37\mathrm{s}\rangle$ state to the $|37\mathrm{p}\rangle$ state on $F_{\mathrm{int}}$. (a) Measurements carried out by direct state-selective electric field ionization of the Rydberg atoms with the NH$_3$ source operated at $P_{\mathrm{NH}_3}=1.8$~mbar (adapted from Ref.~\cite{zhelyazkova17a}). (b) Data recorded by Rydberg-state-selective electric field ionization following microwave transfer from the $|37\mathrm{p}\rangle$ state to the $|37\mathrm{d}\rangle$ state with $P_{\mathrm{NH}_3}=0.9$~mbar.\label{fig6}}
\end{figure}

\subsection{Electric-field controlled F\"orster resonance energy transfer}

To probe changes in the resonant transfer of energy with the strength of the interaction electric field, $F_{\mathrm{int}}$ was varied while monitoring the integrated electron signal in the shaded windows between the ionization fields of~215 and~240~V/cm in Fig.~\ref{fig4}, and~350 and~360~V/cm in Fig.~\ref{fig5}. The corresponding data, recorded by direct electric field ionization and by the microwave assisted detection scheme, are presented in Fig.~\ref{fig6}(a) and (b), respectively. When recording the data in Fig.~\ref{fig6}(a) the NH$_3$ source was operated at 1.8~mbar, while for that in Fig.~\ref{fig6}(b) it was operated at 0.9~mbar. Both of these sets of data exhibit a similar resonance feature at $F_{\mathrm{int}}\simeq5$~V/cm. This is the field for which the resonance condition for the energy transfer process and the electric field dependence of the cross section for resonant energy transfer conspire to give the maximal energy transfer rate under the experimental conditions. The two sets of experimental data in Fig.~\ref{fig6} are compared with the results of calculations in which the constant of proportionality associated with the resonance width, $C_{\mathrm{width}}$ in Eq.~\ref{eq:width} (see Sec.~\ref{sec:theory}), was set to~18 and 23~D\,cm$^{-1}$ ($\equiv18\times10^{-19}$ and $\equiv23\times10^{-19}$~C\,m/s) as indicated by the continuous and dashed curves, respectively. 

The data in Fig.~\ref{fig6}(a) recorded at the higher NH$_3$ source pressure by direct electric field ionization follow more closely the calculated curve for which $C_{\mathrm{width}} = 18$~D\,cm$^{-1}$. On resonance, when $F_{\mathrm{int}} = 5$~V/cm and $\mu_{37\mathrm{s}',37\mathrm{p}'} = 2050$~D, the pseudo first-order kinetic model yields a resonance width of 0.33~cm$^{-1}$ ($\equiv9.8$~GHz) for this value of $C_{\mathrm{width}}$. Consequently, the typical atom-molecule interaction time for collisions that lead to observable energy transfer is inferred to be 0.1~ns, and from Eq.~\ref{eq:width} the corresponding relative collision speed is approximately $350$~m/s. On the other hand the measurements carried out with microwave assisted detection are in better agreement with the calculations for which $C_{\mathrm{width}} = 23$~D\,cm$^{-1}$. In this case the resonance width is 0.42~cm$^{-1}$ ($\equiv12.6$~GHz), the typical atom-molecule interaction time is 0.08~ns, and the corresponding relative collision speed is approximately 430~m/s. These differences in the typical relative collision speeds in these two sets of data are attributed in part to the stricter conditions imposed on the longitudinal speed of the component of the Rydberg atom beam that was detected using the microwave assisted detection scheme, and deviations in the relative directions of propagation of the He and NH$_3$ beams in the two experiments. However, they may also be indicative of a departure from the simple theoretical model based on pseudo first order kinetics. More generally these results highlight the sensitivity of the measurements to the relative collision speed of the atoms and molecules and provide a strong motivation for performing future experiments with velocity controlled beams. 

\begin{figure}
\includegraphics[clip=,width=7.5cm]{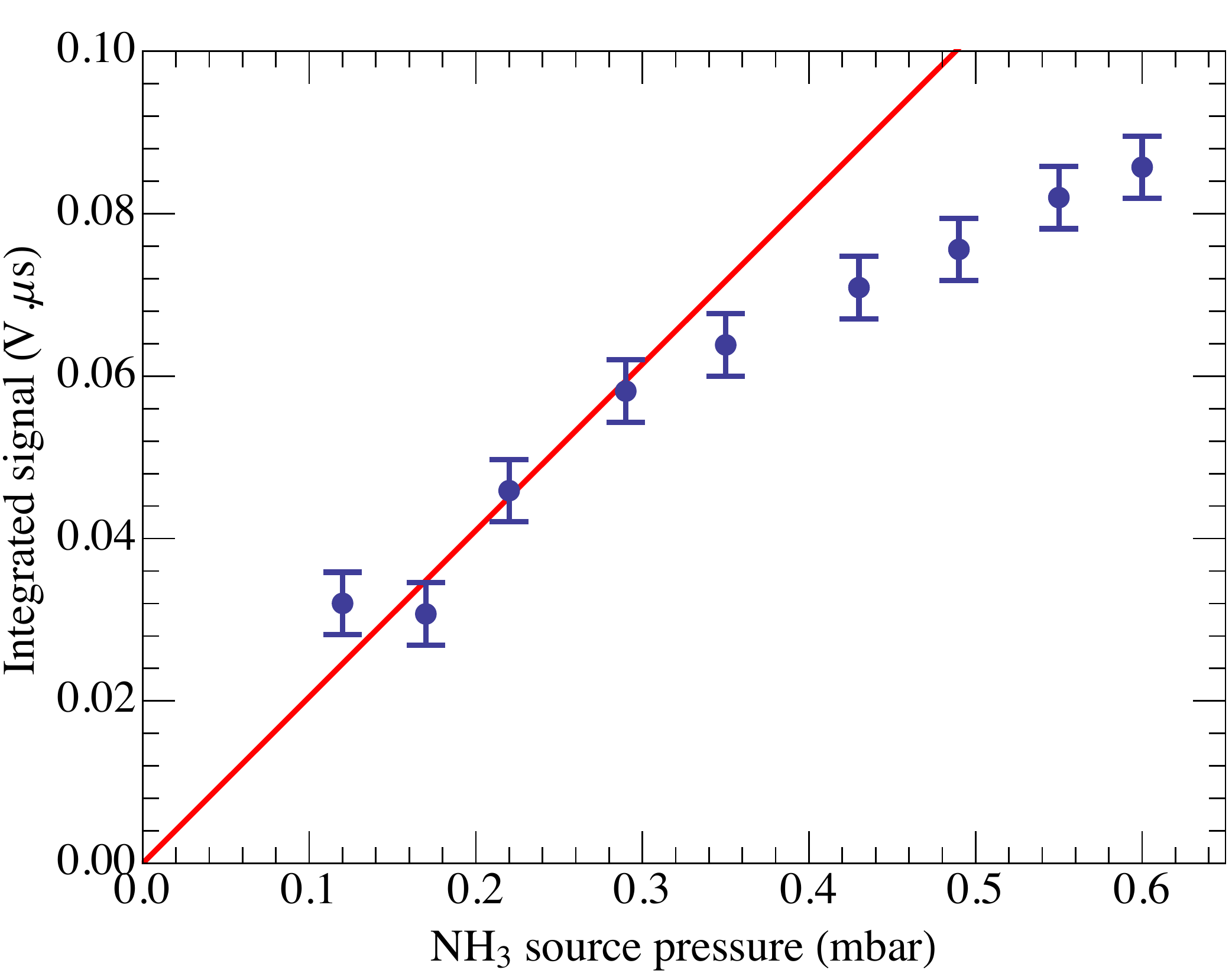}
\caption{The NH$_3$-source-pressure dependence of the integrated $|40\mathrm{d}\rangle$ electron signal following microwave transfer from the $|40\mathrm{p}\rangle$ state in atoms having undergone resonant energy transfer. The experiments were performed in zero electric field. The continuous line represents a least-squares fit to the experimental data for $P_{\mathrm{NH}_3} < 0.4$~mbar.\label{fig7}}
\end{figure}

\subsection{Effects of molecular beam density} \label{sec:dens}

The theoretical model presented in Sec.~\ref{sec:theory} and used to describe the resonant energy transfer process is based on pseudo first order chemical kinetics. For this model to hold, it therefore requires that (1) there is an excess of NH$_3$ in the collision environment so that the NH$_3$ density does not change significantly within a single experimental cycle, (2) the energy transfer channel isolated for study is not coupled to other channels, i.e., in a collision in which energy is transferred from the inversion sublevels of NH$_3$ to the Rydberg He atoms the molecule does not also undergo rotational or vibrational state changes, or Penning ionization, and (3) the atoms to which energy is transferred in the collisions with the NH$_3$ experience no more than one inelastic collision. In the effusive molecular beams used in the experiments these conditions are only expected to hold for low NH$_3$ densities. At higher densities, the large number of rotational states populated are expected to result in significant contributions particularly from processes (2) and (3) above. 

To identify the pressures at which the NH$_3$ beam can be operated while remaining in a regime in which this pseudo first order kinetic model is valid, measurements were made using the microwave assisted detection scheme with atoms initially excited to the $|40\mathrm{s}\rangle$ state. In recording this data $F_{\mathrm{int}}=0$~V/cm, and for each repetition of the experiment $P_{\mathrm{NH}_3}$ was adjusted. The results are displayed in Fig.~\ref{fig7}. For the lower NH$_3$ source pressures used in recording these data, i.e., those below $\sim0.4$~mbar, the integrated signal resulting from the energy transfer process increases approximately linearly with source pressure. This is expected for a first order process. However, for pressures above $\sim0.4$~mbar the rate at which the energy transfer signal changes reduces. This is indicative of a change from a first order process to one in which multiple collisions, or multiple collision channels, start to play a role. The results of these measurements suggest that for the experiments described here and in Ref.~\cite{zhelyazkova17a}, which were performed at NH$_3$ source pressures close to and above 1~mbar with NH$_3$ densities on the order of $\sim10^{9}$~cm$^{-3}$, the first order kinetic model only provides an approximate description of the dynamics. The use of quantum-state-selected molecular beams in future experiments will provide opportunities to accurately probe the broad range of competing processes that contribute to the transition away from regime in which pseudo first order kinetics dominate.

\subsection{Microwave spectroscopy} \label{sec:dens}

\begin{figure}
\includegraphics[clip=,width=8.5cm]{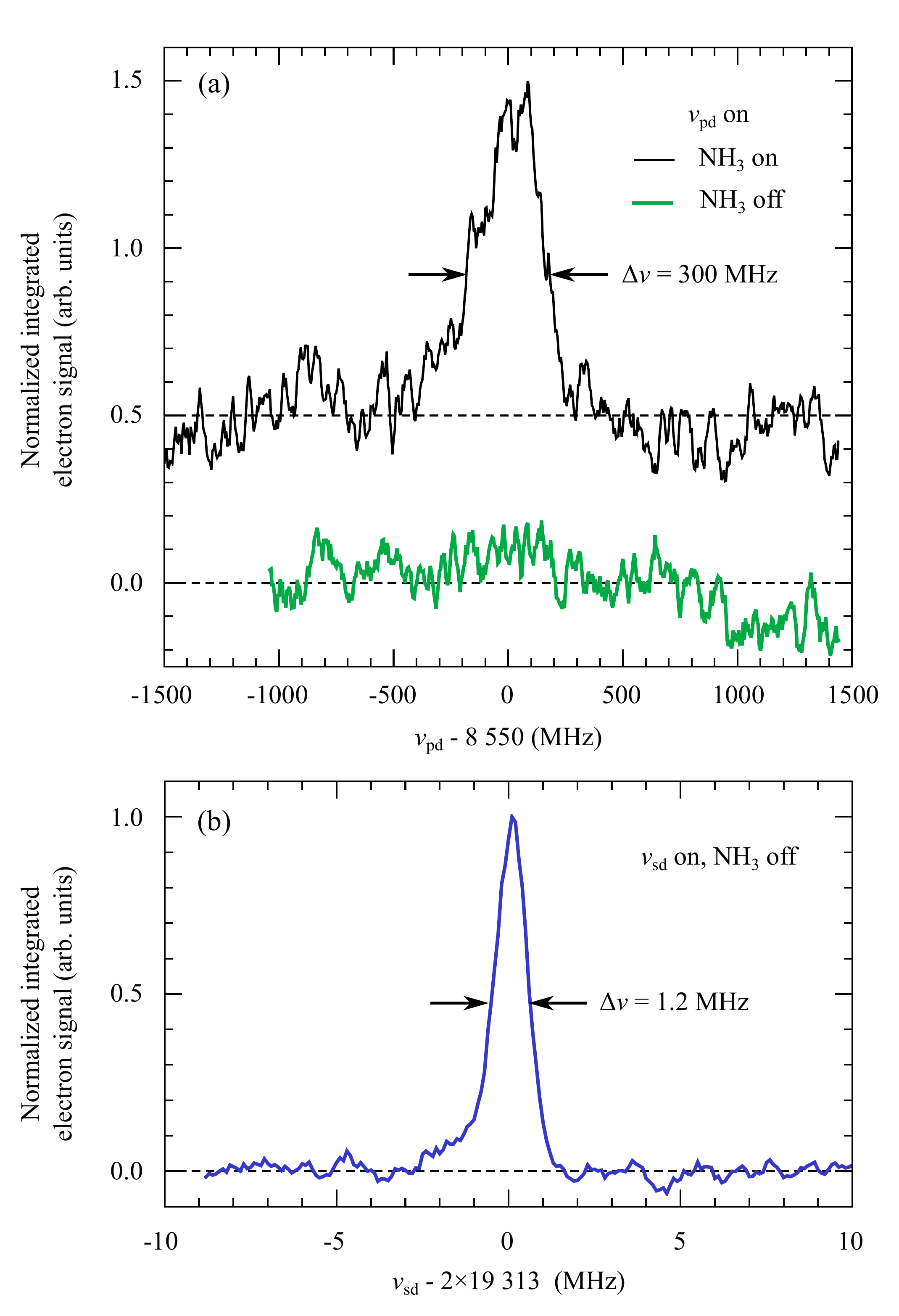}
\caption{\label{fig8} Microwave spectra of (a) the $|37\mathrm{p}\rangle\rightarrow|37\mathrm{d}\rangle$ transition with (upper black spectrum) and without (lower green spectrum) NH$_3$ present, and (b) the two-photon $|37\mathrm{s}\rangle\rightarrow|37\mathrm{d}\rangle$ transition recorded using the same integration windows as in (a) to characterize stray electric fields in the interaction region in the absence of NH$_3$. The upper spectrum in (a) is vertically offset by 0.5 for clarity.}
\end{figure}

Further evidence of effects arising from collision channels other than those associated with the resonant energy transfer channel of interest have been identified by microwave spectroscopy of the $|37\mathrm{p}\rangle\rightarrow|37\mathrm{d}\rangle$ transition in atoms that have undergone resonant energy transfer. Such spectra are displayed in Fig.~\ref{fig8}(a). The upper spectrum, vertically offset by 0.5 in this figure, was recorded with the NH$_3$ beam on while the lower one was recorded in the absence of NH$_3$. Without NH$_3$ the collisional transfer of population from the initially laser photoexcited $|37\mathrm{s}\rangle$ state to the $|37\mathrm{p}\rangle$ state, when $F_{\mathrm{int}}=4.9$~V/cm as in Fig.~\ref{fig5}, does not occur. Consequently, when the microwave spectrum was recorded no resonance was observed. On the other hand, when the $|37\mathrm{p}\rangle$ state was populated through resonant energy transfer from the NH$_3$, a clear spectral feature, at the frequency of the $|37\mathrm{p}\rangle\rightarrow|37\mathrm{d}\rangle$ transition, $\nu_{\mathrm{pd}}=8.550$~GHz, is seen. 

In recording the spectra in Fig.~\ref{fig8}(a) the microwave pulses used had durations of 2~$\mu$s. Hence, Fourier-transform-limited spectra are expected to exhibit resonances with widths of $\sim0.5$~MHz. However, the spectral feature in the upper part of this figure has a width of $\sim300$~MHz. The spectral broadening is attributed to the presence of ions in the interaction region of the apparatus when the NH$_3$ was present. To shift the $|37\mathrm{p}\rangle\rightarrow|37\mathrm{d}\rangle$ transition by half of the measured spectral width, i.e., 150 MHz, a field of 150~mV/cm would be required. This corresponds to the electric field at a distance of 10~$\mu$m from a singly-charged cation, and suggests the presence of a significant number of ions in the interaction region, possibly at number densities up to $10^8$~cm$^{-3}$, when the NH$_3$ beam is present. Such high densities of ions are not present in the apparatus when the NH$_3$ beam is off. This is confirmed in the two-photon spectrum of the $|37\mathrm{s}\rangle\rightarrow|37\mathrm{d}\rangle$ transition in Fig.~\ref{fig8}(b) recorded using the same integration window as the spectra in (a). The shift of this measured two-photon transition frequency by less than 300~kHz from that predicted from the quantum defects in Ref.~\cite{drake99} suggests that in the absence of NH$_3$, stray electric fields in the interaction region of the apparatus are cancelled to $\sim7$~mV/cm. The ions in the interaction region when the NH$_3$ is present result from a combination of (1) Penning ionization of NH$_3$ in collisions with He atoms in the metastable 1s2s\,$^3$S$_1$ level, (2) Penning ionization of NH$_3$ in collisions with Rydberg He atoms, and (3) ionization of Rydberg He atoms following energy transfer from the rotational degrees of freedom of the thermal NH$_3$ beam. In future experiments contributions from collisional ionization processes to the production of such ions in the interaction region of the apparatus can be reduced by using state-selected supersonic beams of NH$_3$, and by guiding or deflecting the excited Rydberg atoms away from the ground state and metastable components of the atomic beam~\cite{seiler11a,allmendinger14a,lancuba13a} before interacting with the NH$_3$.

\section{Conclusion}\label{sec:conc}

In conclusion, resonant energy transfer from the inversion sublevels in thermal beams of NH$_3$ to He atoms in triplet Rydberg states has been studied in zero applied electric field, and in static electric fields of up to 12~V/cm. By using direct Rydberg-state-selective electric field ionization and microwave spectroscopic methods to isolate individual resonant energy transfer channels, the resonant behaviour of the energy transfer process was demonstrated by tuning the strength of the applied electric fields. The electric field dependence of the energy transfer process is in good agreement with the predictions of a theoretical model. 

Comparison of the results presented here and in Ref.~\cite{zhelyazkova16a}, with earlier studies of $\ell$-changing, in collisions of Rydberg atoms with polar molecules which do not occur through F\"orster resonance~(see, e.g., Ref.~\cite{gallagher92a} and references therein) but rather as a result of Rydberg-electron scattering from the polar molecule perturber within its orbit, suggests that for the isolated energy transfer channels studied here at least 80\% of the detected energy-transfer signal results from the F\"orster resonance process. This bound is limited by the uncertainties in the experimental data. It will be of interest, and importance, in future experiments to further distinguish the roles that each of these Rydberg-atom--molecule interactions play in the resonant energy transfer process, and identify contributions from these and other long-range interactions between the collision partners, to the conditions for the resonant transfer of energy.

In zero applied electric field the evolution of the energy transfer process from one described by pseudo first order chemical kinetics to one in which multiple collisions, or multiple collision channels, begin to play a role was identified as the NH$_3$ density was increased. By performing microwave spectroscopy of the Rydberg atoms after undergoing resonant energy transfer, the presence of stray ions in the interaction region of the apparatus could be identified. Effects of these ions on the energy transfer process will be controlled and the measured resonance widths reduced in future experiments through the use of quantum-state-selected supersonic molecular beams, and inhomogeneous electric fields to merge these with the Rydberg atom beams~\cite{allmendinger16a,allmendinger16b}. This will open up new opportunities for studies of molecular dynamics at low temperatures and low collision energies in which long-range electric dipole interactions can be exploited to regulate access to short range processes including Penning ionization, and ion-molecule chemistry.

\begin{acknowledgments}
This work is supported by the Engineering and Physical Sciences Research Council under Grant No. EP/L019620/1, and the European Research Council (ERC) under the European Union's Horizon 2020 research and innovation programme (grant agreement No 683341).
\end{acknowledgments}


\begin{thebibliography}{99}

\bibitem{forster48} T. F\"{o}rster, Ann. Physik. {\bf 437}, 55 (1948).
\bibitem{scholes03a} G. D. Scholes, Annu. Rev. Phys. Chem. {\bf 54}, 57 (2003).
\bibitem{latt65} S. Latt, H. T. Cheung, E. R. Blout, J. Am. Chem. Soc. {\bf 87}, 995 (1965). 
\bibitem{stryer78} L. Stryer, Annu. Rev. Biochem. {\bf 47}, 819 (1978).
\bibitem{millar01} D. Kostermeier and D. P. Millar, Methods {\bf 23}, 240 (2001). 
\bibitem{anderson49} P. W. Anderson, Phys. Rev. {\bf 76}, 647 (1949). 
\bibitem{heijmen99} T. G. A. Heijmen, R. Moszynski, P. E. S. Wormer, A. van der Avoird, A. D. Rudert, J. B. Halpern, J. Martin, W. Bin Gao and H. Zacharias, J. Chem. Phys {\bf 111}, 2519 (1999).

\bibitem{hotop70a} H. Hotop and A. Niehaus, Int. J. Mass Spectrom. Ion Phys. {\bf 5}, 415 (1970).
\bibitem{henson12a} A. B. Henson, S. Gersten, Y. Shagam, J. Narevicius, and E. Narevicius, Science  {\bf 338}, 234 (2012).
\bibitem{jankunas15a} J. Jankunas, K. Jachymski, M. Hapka, and A. Osterwalder, J. Chem. Phys. {\bf 142}, 164305 (2015).

\bibitem{zeppenfeld17a} M. Zeppenfeld, EPL {\bf 118}, 13002 (2017).
\bibitem{kuznetsova16a} E. Kuznetsova, S. T. Rittenhouse, H. R. Sadeghpour, and S. F. Yelin,
Phys. Rev. A {\bf 94}, 032325 (2016).
\bibitem{kuznetsova11a} E. Kuznetsova, S. T. Rittenhouse, H. R. Sadeghpour, and S. F. Yelin, Phys. Chem. Chem. Phys., {\bf 13}, 17115 (2011).
\bibitem{huber12} S. D. Huber and H. P. B\"{u}chler, Phys. Rev. Lett. {\bf 108}, 193006 (2012).
\bibitem{zhao12} B. Zhao, A. W. Glaetzle, G. Pupillo, and P. Zoller, Phys. Rev. Lett. {\bf 108}, 193007 (2012).

\bibitem{safinya81} K. A. Safinya, J. F. Delpech, F. Gounand, W. Sandner, and T. F. Gallagher, Phys. Rev. Lett. {\bf 47}, 405 (1981).  
\bibitem{gallagher92a} T. F. Gallagher, Phys. Rep. {\bf 210}, 319 (1992).
\bibitem{vogt06a} T. Vogt, M. Viteau, J. Zhao, A. Chotia, D. Comparat, and P. Pillet, Phys. Rev. Lett. {\bf 97}, 083003 (2006). 
\bibitem{gallagher08a} T. F. Gallagher and P. Pillet, Adv. At. Mol. Opt. Phys. {\bf 56}, 161 (2008).
\bibitem{gunter13a} G. G\"unter, H. Schempp, M. Robert-de-Saint-Vincent, V. Gavryusev, S. Helmrich, C. S. Hofmann, S. Whitlock, and M. Weidem\"uller, Science {\bf 342}, 954 (2013).
\bibitem{ravets14} S. Ravets, H. Labuhn, D. Barredo, L. B\'eguin, T. Lahaye. and A. Browaeys, Nature Phys. {\bf 10}, 914 (2014).

\bibitem{smith78} K. A. Smith, F. G. Kellert, R. D. Rundel, F. B. Dunning, and R. F. Stebbings, Phys. Rev. Lett. {\bf 40}, 1362 (1978).
\bibitem{petitjean86} L. Petitjean, F. Gounand, and P. R. Fournier, Phys. Rev. A {\bf 33}, 143 (1986).
\bibitem{petitjean84} L. Petitjean, F. Gounand, and P. R. Fournier, Phys. Rev. A {\bf 31}, 71 (1984).
\bibitem{ling93} X. Ling, M. T. Frey, K. A. Smith, and F. B. Dunning, J. Chem. Phys. {\bf 98}, 2486 (1993).

\bibitem{zhelyazkova17a} V. Zhelyazkova and S. D. Hogan, Phys. Rev. A {\bf 95}, 042710 (2017). 

\bibitem{townes55a} C. H. Townes and A. L. Schawlow, \emph{Microwave Spectroscopy} (MCGraw-Hill, London, 1955).

\bibitem{zimmerman79a} M. L. Zimmerman, M. G. Littman, M. M. Kash, and D. Kleppner, Phys. Rev. A {\bf 20}, 2251 (1979).

\bibitem{drake99} G. W. F. Drake, Phys. Scr. {\bf T83}, 83 (1999).

\bibitem{halfmann00} T. Halfmann, J. Koensgen, and K. Bergmann, Meas. Sci. Technol. {\bf{11}}, 1510 (2000).

\bibitem{zhelyazkova16a} V. Zhelyazkova, R. Jirschik, and S. D. Hogan, Phys. Rev. A {\bf 94}, 053418 (2016).

\bibitem{scholes88a} G. Scholes, \emph{Atomic and Molecular Beam Methods, vol. 1}, (Oxford University Press, New York, 1988).

\bibitem{seiler11a} Ch. Seiler, S. D. Hogan, H. Schmutz, J. A. Agner and F. Merkt, Phys. Rev. Lett. {\bf 106}, 073003 (2011).

\bibitem{lancuba13a} P. Lancuba and S. D. Hogan, Phys. Rev. A {\bf 88}, 043427 (2013).

\bibitem{allmendinger14a} P. Allmendinger, J. Deiglmayr, J. A. Agner, H. Schmutz, and F. Merkt, Phys. Rev. A {\bf 90}, 043403 (2014).

\bibitem{allmendinger16a} P. Allmendinger, J. Deiglmayr, K. H{\"o}veler, O. Schullian, and F. Merkt, J. Chem. Phys. {\bf 145}, 244316 (2016).

\bibitem{allmendinger16b} P. Allmendinger, J. Deiglmayr, O. Schullian, K. H{\"o}veler, J. A. Agner, H. Schmutz, and F. Merkt, Phys. Chem. Chem. Phys. {\bf 17}, 3596 (2016).


\end{thebibliography}
\end{document}